\documentclass[smallextended]{svjour3}       
\smartqed  
\usepackage{graphicx}
\usepackage[cmex10]{amsmath}
\usepackage{amssymb}
\usepackage{amsbsy}
\usepackage{algorithmicx,algorithm}
\usepackage{algpseudocode}
\usepackage{threeparttable}
\usepackage{enumerate}
 \usepackage{mathptmx}      
\newtheorem{thm1}{Theorem}

\newtheorem{lma1}{Lemma}
\newtheorem{lma2}[lma1]{Lemma}

\newtheorem{cor1}{Corollary}
\newtheorem{cor2}[cor1]{Corollary}

\newtheorem{exp1}{Counterexample}
\newtheorem{exp2}[exp1]{Counterexample}

\begin{document}

\title{On a Theorem of Kyureghyan and Pott
\thanks{Communicated to and by Pascale Charpin, Alexander Pott}
}


\author{Minglong Qi \and Shengwu Xiong  }


\institute{Minglong Qi \and Shengwu Xiong  \at
              School of Computer Science and Technology,  Wuhan University of Technology \\
              Mafangshan West Campus, 430070 Wuhan City, China\\
              \email{mlqiecully@163.com (Minglong Qi) }\\
              \email{xiongsw@whut.edu.cn (Shengwu Xiong)}\\ 
}

\date{Received: date / Accepted: date}

\maketitle

\begin{abstract}
In the paper of Gohar M. Kyureghyan and Alexander Pott (Designs, Codes and Cryptography, 29, 149-164, 2003), the linear feedback polynomials of the
Sidel\textquoteright nikov-Lempel-Cohn-Eastman sequences were determined for some special cases. When referring to that paper, we found that Corollary 4 and Theorem 2 of that paper are wrong because there exist many counterexamples for these two results. In this note, we give some counterexamples of Corollary 4 and Theorem 2 of that paper.

\end{abstract}

\section{Introduction}

Let $ q $ be a prime power, $ \mathbb{F}_{q} $ be the finite field with $ q $ elements, and $ \mathbb{F}_{q}^{*}= \mathbb{F}_{q}\setminus \lbrace 0\rbrace$. Let $ S=(s_{0},s_{1},s_{2},\cdots) $ be a sequence each term of which is taken from  $ \mathbb{F}_{q} $. Let $ N $ be a positive integer. The sequence $ S $ is said to be $ N- $ periodic 
if $ s_{i+N} =s_{i}$ for all $ i\geq 0 $. The $ N- $periodic sequence $ S $ is denoted by $ S_{N}=(s_{0},s_{1},s_{2},\cdots,s_{N-1}) $. Define $ S_{N}(x)\in \mathbb{F}_{q}[x] $ to be the polynomial 
\begin{equation*}
S_{N}(x)=s_{0}+s_{1}x+s_{2}x^{2}+\cdots+s_{N-1}x^{N-1}.
\end{equation*}
The linear complexity of $ S_{N} $ is defined to be the smallest positive integer, $ L $, such that there exist $ c_{0}=1,c_{1},\cdots,c_{L}\in \mathbb{F}_{q} $ satisfying
\begin{equation*}
-a_{i}=c_{1}a_{i-1}+c_{2}a_{i-2}+\cdots+c_{L}a_{i-L}\ \text{for all}\ L\leq i.
\end{equation*}
It is clear that the linear complexity, $ L $, of the sequence $ S_{N}  $, is the length of the shortest linear feedback register which generates the sequence. The polynomial 
\begin{equation*}
c(x)=c_{0}+c_{1}x+c_{2}x^{2}+\cdots+c_{L}x^{L} 
\end{equation*}
is refferred to as the linear feedback polynomial of the shortest linear feedback shift register that generates $ S_{N}(x) $. It is well kown \cite{bib01,bib02},  that  the linear feedback polynomial can be computed by 
\begin{equation}\label{feedback_polynomial}
c(x)=\dfrac{x^{N}-1}{\gcd(S_{N}(x),x^{N}-1)}.
\end{equation}
Hence, the linear complexity can be determined by
\begin{equation}\label{linear+complexity}
L=N-\deg(\gcd(S_{N}(x),x^{N}-1)).
\end{equation}

Let $ q=df+1 $, and $ \alpha $ be a primitive element of $ \mathbb{F}_{q} $. The cosets 
\begin{equation*}
C_{i}^{d}=\lbrace \alpha^{kd+i}|0\leq k\leq f-1\rbrace, i=0,\cdots,d-1
\end{equation*}
are called the cyclotomic classes of order $ d $ with respect to $ \mathbb{F}_{q} $. Note that the cyclotomic classes $ C_{i}^{d} $ depend on the choice of the primitive element $ \alpha $. It is clear that  
\begin{equation*}
\mathbb{F}_{q}^{*}=\bigcup_{i=0}^{d-1}C_{i}^{d}.
\end{equation*}
The constants
\begin{equation*}
(l,m)_{d}=\mid\bigl(C_{l}^{d}+1\bigr)\bigcap C_{m}^{d}\mid
\end{equation*}
are called the cyclotomic numbers of order $ d $ with respect to $ \mathbb{F}_{q} $. 

Let $ q=p^{m} $ where $ p $ is an odd prime, and $ m $ a positive integer. If $ p\equiv 1\pmod 4 $, then $ q $ can be represented by the Diophantine equation $ q=x^{2}+4y^{2} $. 
If $ \gcd(x,q)=1,x\equiv 1 \pmod 4 $, the representation is  called the proper representation of $ q $.

The quadratic character of $ \mathbb{F}_{q}^{*} $ is defined by
\begin{equation*}
\eta(\beta)=
\begin{cases}
1 &\qquad\text{if}\ \beta=\gamma^{2}\ \text{for some}\ \gamma\in\mathbb{F}_{q}^{*}\\
0 &\qquad\text{if}\ \beta=0\\
-1 &\qquad\text{otherwise}.
\end{cases}
\end{equation*}
Let $ a\in \mathbb{F}_{q}^{*} $, and $ n\in \mathbb{N} $. Based on the quadratic character of $ \mathbb{F}_{q}^{*} $,  two types of Jacobsthal sums \cite{bib02,bib03} are defined by
\begin{equation}\label{jacobsthal_sums}
\begin{split}
I_{n}(a)&=\sum_{c\in \mathbb{F}_{q}^{*}}\eta(c^{n}+a),\\
H_{n}(a)&=\sum_{c\in \mathbb{F}_{q}^{*}}\eta(c)\eta(c^{n}+a).
\end{split}
\end{equation}

Recall that $ \alpha $ is a primitive element of $ \mathbb{F}_{q} $. The Sidel\textquoteright nikov-Lempel-Cohn-Eastman sequence $ S_{q}=(s_{0},s_{1},\cdots,s_{q-2}) $ of period $ q-1 $ over $ \mathbb{F}_{2} $ is defined by
\begin{equation}\label{SLCE_sequence}
s_{i}=
\begin{cases}
1 &\qquad\text{if}\ \eta(\alpha^{i}+1)=-1\\
0 &\qquad\text{otherwise}.
\end{cases}
\end{equation}

Let $ LC_{q} $ denote the linear complexity of $ S_{q} $ over $ \mathbb{F}_{2} $ and 
\begin{equation}\label{Sq_polynomial}
S_{q}(x)=\sum_{i=0}^{q-2}s_{i}x^{i}\in \mathbb{F}_{2}[x].
\end{equation}
Then, the linear feedback polynomial of $ S_{q} $ is
\begin{equation*}
\dfrac{x^{q-1}+1}{\gcd(x^{q-1}+1,S_{q}(x))}
\end{equation*}
and the linear complexity 
\begin{equation*}
LC_{q}=q-1-\deg(\gcd(x^{q-1}+1,S_{q}(x))).
\end{equation*}

In this paragraph, when we say Corollary 4, Lemma 5, and Theorem 2, we refer to those of  \cite{bib04}. In \cite{bib04}, G. Kyureghyan and A. Pott determined the linear complexity and the linear feedback polynomials of the Sidel\textquoteright nikov-Lempel-Cohn-Eastman sequences for some special cases. When studying the similar problems and referring to that paper, we found that Corollary 4 and Theorem 2  are wrong because there exist many counterexamples. It can be easily seen that the cause making Corollary 4 and Theorem 2  wrong is that, the necessary and sufficient conditions of Corollary 4 and Theorem 2  are not equivalent to the negation of the condition of Lemma 5 , from which Corollary 4 and Theorem 2   follow. The rest of the note is structured as follows: in  second section, some counterexamples of Corollary 4 and Theorem 2   are provided, and the correction of them is given by readopting the negation of the condition of Lemma 5. In section 3, a brief conclusion is given.

\section{Counterexamples of Corollary 4 and Theorem 2 and Correction of the Two Results}
 
Next lemma strengthens an observation in \cite{bib05}.
\begin{lma1}[Lemma 4 in \cite{bib04}]\label{lmma1}
\begin{enumerate}[(a)]
\item 
If $ q\equiv 5\pmod 8 $, then $ x+1 $ divides $ \gcd(x^{q-1}+1,S_{q}(x)) $, and $ (x+1)^{2} $ does not divide $ \gcd(x^{q-1}+1,S_{q}(x)) $ over $ \mathbb{F}_{2} $.
\item If $ q-1=8f\equiv 0\pmod 8 $, then $ (x+1)^{i},i\geq 2 $ divides $ \gcd(x^{q-1}+1,S_{q}(x)) $ over $ \mathbb{F}_{2} $. Moreover, $ (x+1)^{4} $ does not divide $ \gcd(x^{q-1}+1,S_{q}(x)) $ over $ \mathbb{F}_{2} $ if $ f+y/2 $ is odd, where $ y $ is determined from the proper representation of $ q=x^{2}+4y^{2} $.
\end{enumerate}
\end{lma1}

Next lemma is a key one in \cite{bib04}, that gives the necessary and sufficient condition by which the factor $ g(x)=x^{d-1}+x^{d-2}+\cdots+1\in \mathbb{F}_{2}[x] $ divides $ \gcd(x^{q-1}+1,S_{q}(x)) \in \mathbb{F}_{2}[x]$, where $ q=df+1 $:
\begin{lma2}[Lemma 5 in \cite{bib04}]\label{lmma2}
If $ q=df+1 $, $ d $ is odd, $ 4 $ divides $ f $ and $ \alpha $ is a primitive element of $ \mathbb{F}_{q} $, then $ \gcd(x^{q-1}+1,S_{q}(x))\in \mathbb{F}_{2}[x] $ is divisible by
\begin{equation*}
g(x)=x^{d-1}+x^{d-2}+\cdots+1\in \mathbb{F}_{2}[x]
\end{equation*}
if and only if 
\begin{equation*}
I_{d}(1)\equiv -d\pmod 4
\end{equation*}
and 
\begin{equation*}
I_{d}(\alpha^{-t})\equiv 0\pmod 4\ \text{for all}\ 1\leq t\leq d-1.
\end{equation*}
\end{lma2} 

Lemma \ref{lmma2} (Lemma 5 in \cite{bib04}) was rigorously proved. Through the proving process, the authors of \cite{bib04} discovered a new polynomial over $ \mathbb{F}_{2} $, namely,
\begin{equation*}
S_{2}(x)=\sum_{t=0}^{d-1}c_{t}x^{t}\in \mathbb{F}_{2}[x],
\end{equation*}
where
\begin{equation*}
c_{t}=\sum_{k=0}^{f-1}s_{t+kd},0\leq t\leq d-1.
\end{equation*}
The authors of \cite{bib04} deduced the necessary and sufficient condition of Lemma \ref{lmma2} is equivalent to $ S_{2}(x)=0 $.

The negation of the necessary and sufficient condition of Lemma \ref{lmma2} is stated by
\begin{equation}\label{neg_condition_lma2}
\begin{split}
I_{d}(1)&\ncong -d\pmod 4\ \text{or}\\ 
I_{d}(\beta)&\ncong 0\pmod 4\ \text{for some}\ \beta\in \lbrace \alpha^{-t}\mid 1\leq t\leq d-1\rbrace.
\end{split}
\end{equation}
Clearly, by Lemma \ref{lmma2}, $g(x)=x^{d-1}+x^{d-2}+\cdots+1\in \mathbb{F}_{2}[x] $ does not divide $ \gcd(x^{q-1}+1,S_{q}(x)) \in \mathbb{F}_{2}[x]$ if only if Eq.(\ref{neg_condition_lma2}) holds. Meanwhile, the authors of \cite{bib04} gave another necessary and sufficient condition, by which $ g(x)\in \mathbb{F}_{2}[x] $ is excluded from being a factor of $ \gcd(x^{q-1}+1,S_{q}(x)) \in \mathbb{F}_{2}[x]$. We will show that that condition is not equivalent to the condition stated in  Eq.(\ref{neg_condition_lma2}). In many counterexamples, both the condition of Corollary 4 in \cite{bib04} and the condition of Lemma \ref{lmma2} hold at the same time, which is absurd. According to the authors of \cite{bib04}, next corollary is an important one of Lemma \ref{lmma2}:

\begin{cor1}[Corollary 4 in \cite{bib04}]\label{cor1_correspond_cor4}
If $ q=2^{k}r+1, k\geq 2, r $ is an odd prime and $ 2 $ is a primitive root modulo $ r $, then
\begin{equation*}
\gcd(x^{q-1}+1,S_{q}(x))=(x+1)^{i}\ \text{for some}\ i\geq 1
\end{equation*}
if and only if 
\begin{equation}\label{cond2_cor4}
\begin{split}
I_{r}(a) &\ncong -r\pmod 4 \ \text{for some}\ a\in \langle \alpha^{r}\rangle\\
&\text{or}\\
I_{r}(a) &\ncong 0\pmod 4 \ \text{for some}\ a\notin \langle \alpha^{r}\rangle,
\end{split}
\end{equation}
where $ \alpha $ is a primitive element of $ \mathbb{F}_{q} $.
\end{cor1}

Remark that the condition of Eq.(\ref{cond2_cor4}) is not always equivalent to that of Eq.(\ref{neg_condition_lma2}). Sometimes, the condition of Lemma \ref{lmma2} and that of Corollary \ref{cor1_correspond_cor4} hold for a same case, which leads to the absurd situation: $ \gcd(x^{q-1}+1,S_{q}(x))\in \mathbb{F}_{2}[x] $ has the factor $ g(x)=x^{r-1}+x^{r-2}+\cdots+1 \in \mathbb{F}_{2}[x]$ because the condition of Lemma \ref{lmma2} is true, and $ \gcd(x^{q-1}+1,S_{q}(x))=(x+1)^{i}\ \text{for some}\ i\geq 1 $ because the condition of Eq.(\ref{cond2_cor4}) hold too. This situation is well illustrated by the following counterexamples:
\begin{exp1}\label{example1}
Let $ q=2^{4}\cdot 3+1=7^{2}, \alpha $ be a primitive element of $ \mathbb{F}_{7^{2}} $. Then,
\begin{equation*}
\begin{split}
&I_{3}(1)=0,\\
&(I_{3}(\beta)+3)\pmod 4 \in \lbrace 0,2\rbrace \ \text{for all }\ \beta\in \langle \alpha^{3}\rangle,\ \text{and}\\
& I_{3}(\beta) \equiv 0 \pmod 4 \ \text{for all }\ \beta\in \mathbb{F}_{q}^{*}\setminus \langle \alpha^{3}\rangle.
\end{split}
\end{equation*}
It means that there are some $ \beta \in \langle \alpha^{3}\rangle $ such that $ (I_{3}(\beta)+3)\equiv 2\pmod 4 $, i.e., $ I_{3}(\beta)\equiv -1 \ncong -3 \pmod 4 $. Hence, the condition of Corollary \ref{cor1_correspond_cor4} is true, and it should be expected that $ \gcd(x^{48}+1,S_{49}(x))=(x+1)^{i} $ for some $ i\geq 1 $. However, $ \gcd(x^{48}+1,S_{49}(x))=(x+1)^{6}(x^{2}+x+1)^{2} $, meaning that Corollary \ref{cor1_correspond_cor4} is wrong. On the other hand, the condition of Lemma \ref{lmma2} is true for this case, which further demonstrates that the condition of Eq.(\ref{cond2_cor4}) is not equivalent to that of Eq.(\ref{neg_condition_lma2}).
\end{exp1}
\begin{exp2}\label{example2}
Let $ q\in\lbrace 193=2^{6}\cdot 3+1,769= 2^{8}\cdot 3+1,12289= 2^{12}\cdot 3+1\rbrace, \alpha $ be a primitive element of $ \mathbb{F}_{q} $. Then,
\begin{equation*}
\begin{split}
&I_{3}(1)=0,\\
&(I_{3}(\beta)+3)\pmod 4 \in \lbrace 0,2\rbrace \ \text{for all }\ \beta\in \langle \alpha^{3}\rangle,\ \text{and}\\
& I_{3}(\beta) \equiv 0 \pmod 4 \ \text{for all }\ \beta\in \mathbb{F}_{q}^{*}\setminus \langle \alpha^{3}\rangle.
\end{split}
\end{equation*}
Clearly, the condition of Corollary \ref{cor1_correspond_cor4} is satisfied. However, $ \gcd(x^{q-1}+1,S_{q}(x))=(x+1)^{2}(x^{2}+x+1)^{2}\neq (x+1)^{i} $ for some $ i\geq 1 $, meaning that Corollary \ref{cor1_correspond_cor4} is wrong. Note that the condition of Lemma \ref{lmma2} holds too, which is absurd.
\end{exp2}

Counterexample \ref{example1}-\ref{example2} show that Corollary \ref{cor1_correspond_cor4} is wrong. This is because the necessary and sufficient condition of Corollary \ref{cor1_correspond_cor4}, stated by Eq.(\ref{cond2_cor4}), is not always equivalent to the negation of the condition of Lemma \ref{lmma2}, expressed by Eq.(\ref{neg_condition_lma2}). The correct version of Corollary \ref{cor1_correspond_cor4} is given by

\begin{cor2}[Correction of Corollary 4 in \cite{bib04}]\label{cor2_correspond_cor4}
If $ q=2^{k}r+1, k\geq 2, r $ is an odd prime and $ 2 $ is a primitive root modulo $ r $, then
\begin{equation*}
\gcd(x^{q-1}+1,S_{q}(x))=(x+1)^{i}\ \text{for some}\ i\geq 1
\end{equation*}
if and only if 
\begin{equation*}
\begin{split}
I_{r}(1)&\ncong -r\pmod 4\ \text{or}\\ 
I_{r}(\beta)&\ncong 0\pmod 4\ \text{for some}\ \beta\in \lbrace \alpha^{-t}\mid 1\leq t\leq d-1\rbrace,
\end{split}
\end{equation*}
where $ \alpha $ is a primitive element of $ \mathbb{F}_{q} $.
\end{cor2}
\begin{proof}
We have 
\begin{equation*}
x^{q-1}+1=x^{2^{k}r}+1=(x^{r}+1)^{2^{k}}=(x+1)^{2^{k}}(x^{r-1}+x^{r-2}+\cdots+1)^{2^{k}}.
\end{equation*}
From the assumption of Corollary \ref{cor2_correspond_cor4}, $ g(x)=x^{r-1}+x^{r-2}+\cdots+1 $ is irreducible over $ \mathbb{F}_{2} $.
By Lemma \ref{lmma2}, $ g(x)=x^{r-1}+x^{r-2}+\cdots+1\in \mathbb{F}_{2}[x] $ is not a factor of $ \gcd(x^{q-1}+1,S_{q}(x))\in \mathbb{F}_{2}[x] $. Hence, $ \gcd(x^{q-1}+1,S_{q}(x)) $ must only have the factor $ (x+1)^{i} $ for some $ 0\leq i\leq 2^{k} $. Note that for $ q-1\equiv 0\pmod 4, x+1 $ is always a factor of $ S_{q}(x) $ \cite{bib05}. Therefore, 
\begin{equation*}
\gcd(x^{q-1}+1,S_{q}(x))=(x+1)^{i}\ \text{where}\ 1\leq i\leq 2^{k}.
\end{equation*} \qed
\end{proof}

Theorem 2 of \cite{bib04} was deduced from Lemma 4  and Corollary 4  of the same paper. Since  Corollary 4 of \cite{bib04} is wrong, Theorem 2 of \cite{bib04} is wrong too. We give the correct version of Theorem 2 of \cite{bib04} by readopting the condition expressed in Eq.(\ref{neg_condition_lma2}) which excludes $ g(x)=x^{r-1}+x^{r-2}+\cdots+1\in \mathbb{F}_{2}[x] $ from being a factor of $ \gcd(x^{q-1}+1,S_{q}(x)) $ by Corollary \ref{cor2_correspond_cor4}:

\begin{thm1}[Correction of Theorem 2 in \cite{bib04}]\label{thm1}
Let $ q=2^{k}r+1,k\geq 1,r $ be an odd prime and $ q=x^{2}+4y^{2} $ be the proper representation of $ q $. If $ 2 $ is a primitive root modulo $ r $, then the feedback polynomial of $ S_{q} $ over $ \mathbb{F}_{2} $ is
\begin{equation*}
\begin{split}
&\dfrac{x^{q-1}+1}{x+1}\ \text{if}\ k=2\ \text{and}\\
&\dfrac{x^{q-1}+1}{(x+1)^{i}}\ \text{for some }\ i\geq 2\ \text{if}\ k\geq 3,
\end{split}
\end{equation*}
(where $ i\leq 4 $ if $ \frac{2^{k-2}r+y}{2} $ is odd) if and only if  
\begin{equation*}
\begin{split}
I_{r}(1)&\ncong -r\pmod 4\ \text{or}\\ 
I_{r}(\beta)&\ncong 0\pmod 4\ \text{for some}\ \beta\in \lbrace \alpha^{-t}\mid 1\leq t\leq r-1\rbrace,
\end{split}
\end{equation*}
where $ \alpha $ is a primitive element of $ \mathbb{F}_{q} $.
\end{thm1}
\begin{proof}
Follows from Lemma \ref{lmma1} and Corollary \ref{cor2_correspond_cor4}. \qed
\end{proof}

\section{Conclusion}
In this note, we show that Corollary 4 and Theorem 2 of \cite{bib04} are wrong by some counterexamples. We point out that the necessary and sufficient condition of Corollary 4 of \cite{bib04} is not equivalent to the negation of the condition of Lemma 5 in \cite{bib04}, which is the cause making Corollary 4 and Theorem 2 of \cite{bib04} wrong. And finally, we correct Corollary 4 and Theorem 2 of \cite{bib04} by readopting the condition stated in Eq.(\ref{neg_condition_lma2}).

\end{document}